# MODELS OF STEP BUNCHING: TURNING REPULSION INTO ATTRACTION

D. Staneva, B. Ranguelov, V. Tonchev[*]
Institute of Physical Chemistry, Bulgarian Academy of Sciences, 1113 Sofia, Bulgaria

**Abstract**. We report numerical results for two models of vicinal motion. The first, LW, aims at crystal evaporation when the detachment from steps is slow [Liu and Weeks, PRB **57**, 23 (1998) 14891]. The source of destabilization is electromigration force acting on the adatoms. The destabilizing part of equation(s) of the step velocity is linear in the widths of the adjacent terraces with larger contribution of the terrace behind. This asymmetry is controlled by a single parameter $b$. The stabilization part accounts for the tendency to equidistant spacing dictated by the interstep repulsions. We construct the second model, LW2, from LW in the same manner as was constructed Minimal Model 2 (MM2) from another minimal model [B.Ranguelov et al., Nanoscience and Nanotechnology 6, (2006) 31] - keeping the 'repulsions term' from LW and introducing a similar one with opposite sign as 'attractions term'. For LW we obtain for first time that in the pre-factor of the time scaling of the number of steps in the bunch $N$ enters only the parameter $b$. In LW2 we find the same type of step bunching as in MM2 - the surface slope in the bunch is constant and not function of $N$. Further, we obtain the time scaling of $N$ with exponent ~0.18, found also in experiments, and the minimal distance in the bunch as function of the model parameters.

**Introduction**. The early work on the step bunching was focused on identifying the destabilizing factors as the Ehrlich-Schwoebel effect [1] and electromigration of the adatoms [2] as well as tracing the theoretical approach to the loss of stability [3]. After the experimental observation of the phenomenon on the Si(111)-vicinal surface[4] the field of surface instabilities undergoes explosive growth and now it is in a stage of maturity. The activity was focused initially on modeling details of the loss of stability but later also the long times of the bunching process become subject of intensive studies [5,6,7,8] combining numerical calculations on discrete models and analytical treatment of continuum ones. Specifically, the size scaling of the minimal bunch distance with the number of steps in the bunch $N$, $l_{min} \sim N^{-2/3}$, was found in two limiting regimes – when the diffusion on the terraces is slow [8,7] and when the attachment/detachment to/from steps is slow [5]. Based on the cumulated knowledge of scaling in different contexts the idea of universality classes in bunching was developed [5, 9] based on a generalized continuum equations for the time evolution of the crystal surface. The introduction of minimal models [10] is an attempt to find a discrete analogue and reference point to the investigations of universality classes. The studies in this paper are a step further in building the hierarchy of models and identifying their universality classes.

**The Models.** The first model we study, LW, is introduced by Liu and Weeks[6] and studied later [5, 11, 12] in detail. The equation for step velocity of the $i$-th step in a step train:

$$\frac{dx_i}{dt} = \frac{1-b}{2}(x_{i+1} - x_i) + \frac{1+b}{2}(x_i - x_{i-1}) + U(2f_i - f_{i-1} - f_{i+1}) \qquad (1)$$

$x_i$ being the position of the $i$-th step and $b$ introduces asymmetry in the contribution to the step motion by the two adjacent terraces, their widths defined by differences of the type $\Delta x_i = x_{i+1} - x_i$ and thus only the distances between steps that are nearest neighbors are taken into account. The step motion is unstable always when $b$ is positive. The third term on the rhs accounts for the step-step repulsion, $U$ is the rescaled strength of the step-step repulsions with energy ~(distance)$^{-n}$, and $f_i$ is:

$$f_i = \left(\frac{l}{\Delta x_i}\right)^{n+1} - \left(\frac{l}{\Delta x_{i+1}}\right)^{n+1} \quad (2)$$

where $l$ is the interstep distance when all steps are equidistant. This equation contains two opposite tendencies, destabilizing and stabilizing, and the time evolution of step configuration reflects in a complex manner the balance between these.

Applying the same procedure as in previous paper from this series [10] we construct from LW a new model, LW2, defined through the equation of step motion:

$$\frac{dx_i}{dt} = -K(2g_i - g_{i-1} - g_{i+1}) + U(2f_i - f_{i-1} - f_{i+1}) \quad (3)$$

with $K$ – the generalized strength of the interstep attractions, and $g_i$ contains a different from $n$ exponent $p$:

$$g_i = \left(\frac{l}{\Delta x_i}\right)^{p+1} - \left(\frac{l}{\Delta x_{i+1}}\right)^{p+1} \quad (4)$$

**Numerical procedure.** We study both models using the same 'template' consisting of numerical integration of the equations of step motion (fourth order Runge-Kutta) to obtain step positions in sequential moments combined with monitoring schemes [13] to follow different characteristics of the step bunches and their relations. Some details for the monitoring schemes (MS) are given also here.

Monitoring Scheme I (MS-I) - at every time step is counted the number of step bunches, using proper 'bunch definition' for the distance between two steps that are nearest neighbors, in this study always when it is less than the vicinal one $l$. Then we find the average number of steps in the bunch as the sum of all steps in bunches divided by the number of bunches. The average width of the bunch is the sum of all distances between steps (nearest neighbors) in bunches divided by the number of bunches and the average terrace width – the sum of all nearest neighbor distances outside bunches divided by the number of bunches, etc.

Monitoring Scheme II (MS-II) - For every bunch size (number of steps in a bunch), from 2 to the maximal number that may appear during whole bunching process (in every moment of time), we cumulate characteristics for every size: minimal distance $l_{min}$, bunch width, etc. We also plot some qualitative characteristics of the bunching - the step trajectories, surface profile and surface slope. Then one finds the size scaling of the minimal interstep distance in the bunch, and time scaling of the number of steps. In order to distinguish between different models and regimes one should use both qualitative and quantitative criteria – information from the surface morphology and size- and time- scaling, respectively.

**Results.** LW is one of the most studied in the field of step bunching [5, 6, 11, 12]. Therefore, the first three figures do not bear original results but rather are illustrative and build the context for the other results. We show results for the special case of $b = -1$ in which the model becomes 'one-sided' – only the terrace behind a step contributes to its motion. In Figure 1 and Figure 2 are shown the trajectories of the steps and the surface slope, respectively. It is seen from Figure 2 that the minimal interstep distance in the bunch $l_{min}$ appears in the middle of the bunch being function of the number of steps in the bunch $N$, Figure 3 and Figure 4. The exponent from the interstep repulsion law $n$ enters the exponent in the size-scaling relation. For first time this type of size-scaling was obtained in [7] and for LW in [5]. The data for $l_{min}$ is sensitive to the values of $U$, Figure 3, but all data for the time evolution of the bunch size $N$ lie on a universal line in log-log plot with slope 1/2, Figure 5. This exponent was obtained in numerous experimental and theoretical studies. The original result from our study is for the pre-factor in this time-scaling - it includes only the (destabilizing) parameter $b$ but not the stabilizing one $U$. Further detailed study of the numerical pre-factor revealed intervals in the values of $b$ where it is constant. In example, for values of $b$ above 2 and for values of $U < b$, the numerical pre-factor remains constant and equal to $(2b)^{1/2}$. The regime $b > 2$ and $U > b$ needs more studies but based on the available data, we can state that the bunching process becomes quite irregular with respect to the scaling and it is impossible to fit data to straight line in log-log plot. Other region of constant value is when $b = 0.1 \div 0.3$ – there the pre-factor is $b^{1/2}$. Between these two intervals the pre-factor changes in a stepwise manner and this makes hard to find the total dependence but always it is independent of the values of $U$.

<u>LW2.</u> The study of a new model usually starts with linear stability analysis to identify the values of the parameters leading to bunch formation. We leave this treatment for a further publication and study the model numerically, checking directly with calculations whether a given combination of parameters causes bunching. For the choice of values of the exponent $p$ we used as reference study the work of Zuo et al.[14] where $p = 0.5 \div 1.5$, $n$ is usually 2[15] but we used also $n = 3$ to make the study systematic in $n$.

The results we obtain for LW2 characterize thoroughly the model with respect to its universality. On qualitative level, it generates the unusual type of step bunching, observed already when studying MM2 [10] – the interstep distance in the bunch is not function of the number of steps in the bunch, see Figure 6Figure 77 and 8. The constant values of the minimal interstep distance in the bunch, obtained in series of calculations with different values of the parameters, are used to plot the dependence on Figure 9, note that the scaling variable (combination of model parameters) $(U/K)^{1(n-p)}$ is similar to the one in LW. From this scaling one would expect two regimes of parameters, leading to instability: (i) $K > U$ and $n > p$ (the one we use in this study) and (ii) $K > U$ and $n > p$.

Further we find, Figure 10, the value of the time-scaling exponent $\beta \sim 0.18$, as found earlier both in experiments and in a Monte Carlo simulation by Sudoh [16]. The exact value of the exponent and its origins could be found in more detailed studies. What we can summarize based on our present studies is that $\beta$ is not larger than 0.2 and thus can be well distinguished from $\beta \approx 1/3$ obtained for MM2.

**Acknowledgements.** Most of the calculations were done on a computer cluster built with financial support from grants F-1413/2004, BM9/2006 and TK-X-1713/2007 from the Bulgarian National Science Fund and VIRT/NANOPHEN-FP6-INCO-CT-2005-016696. BR and VT acknowledge the support of IRC-CoSim Project. VT acknowledges the hospitality and the stimulating working conditions at Math Department of the University of Kentucky, Lexington, KY, USA.

*Corresponding author: tonchev@ipc.bas.bg

**Figure 1**          **Figure 2**

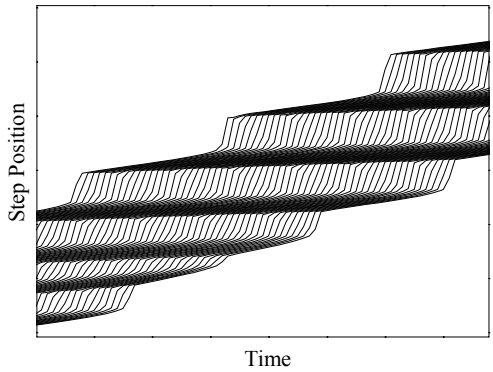 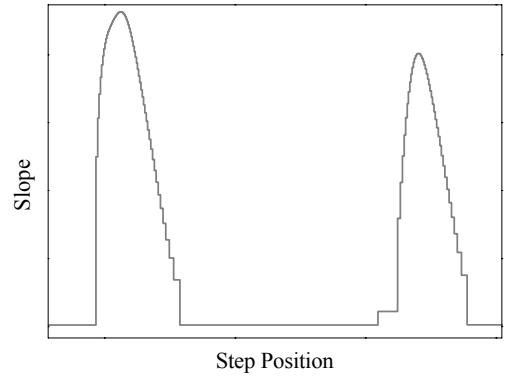

Figure 1 LW, step trajectories obtained integrating numerically equations of step motion, late times. Used values of the parameters: $b = 1$, $U = 0.5$, $n = 2$.

Figure 2 LW, surface slope (inverse of the interstep distance). It is clearly seen that the slope is largest in the middle of the bunch. $b = 1$, $U = 0.5$, $n = 2$

**Figure 3**          **Figure 4**

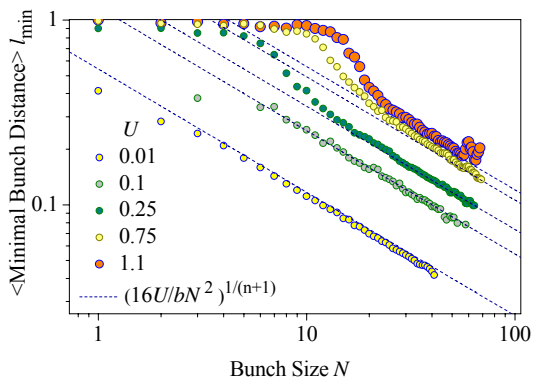 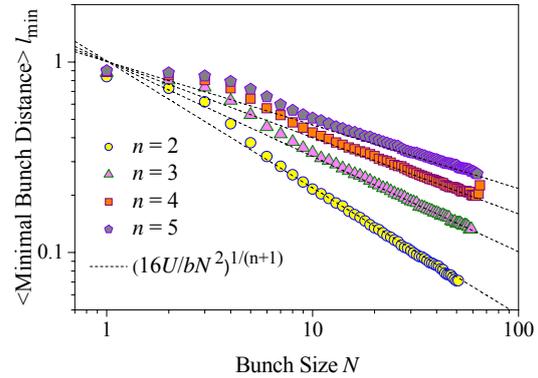

Figure 3 LW, size scaling of the minimal interstep distance in the bunch from MS-II. With increasing the strength of repulsion $U$, the values of for larger bunch sizes. $l_{min}$ fits the prediction [5] becoming less then 0.15-0.2 the initial vicinal distance, $n = 2$, $b = 1$

Figure 4 LW, size scaling of the minimal interstep distance in the bunch for different values of $n$. $U = 0.065$, $b = 1$.

**Figure 5**

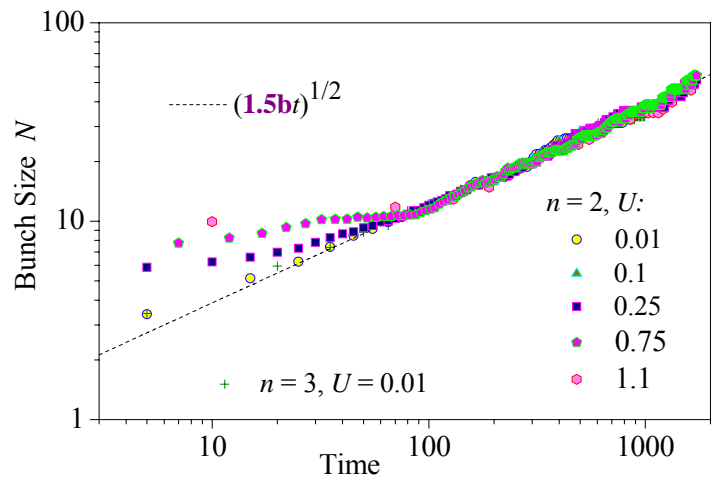

Figure 5 LW, time scaling of the bunch size from Monitoring Scheme - I. For $b = 1$, the numerical pre-factor is $(1.5b)^{1/2}$ independent of $U$ and $n$.

**Figure 6**                                                                                              **Figure 7**

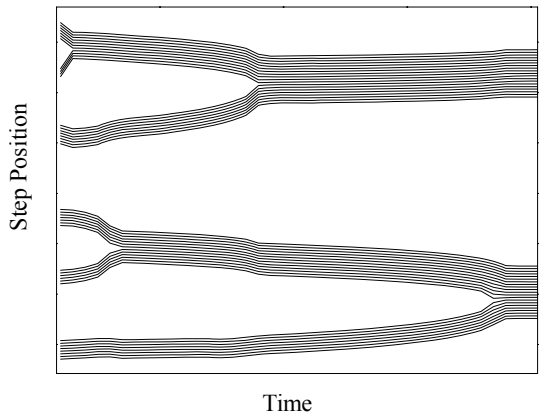 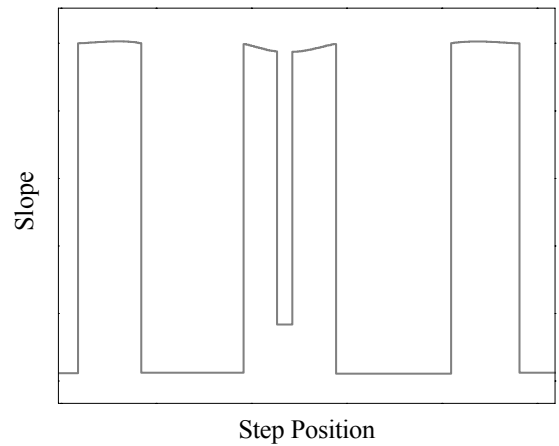

Figure 6 LW2, step trajectories obtained integrating numerically equations of step motion of LW2, early times. Note that the bunching occurs also when the power of interstep attraction $p$ is 0 as in this run. As seen, the steps only reorganize into groups without translation.

Figure 7. LW2, surface slope (inverse of the interstep distance) as a result of the bunching process in LW2. The two bunches in the middle have opposite slopes in the process of their coalescence. The surface between these bunches is steeper then the large terraces separating bunches, $p = 1$

**Figure 8**

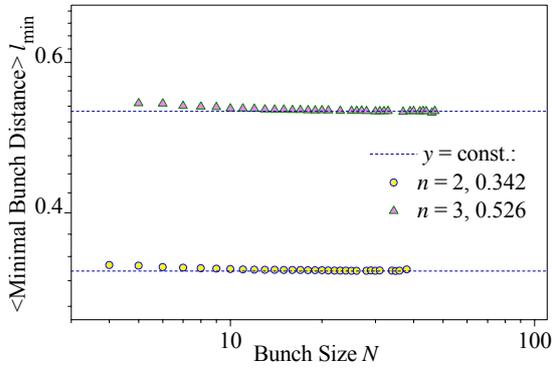

**Figure 9**

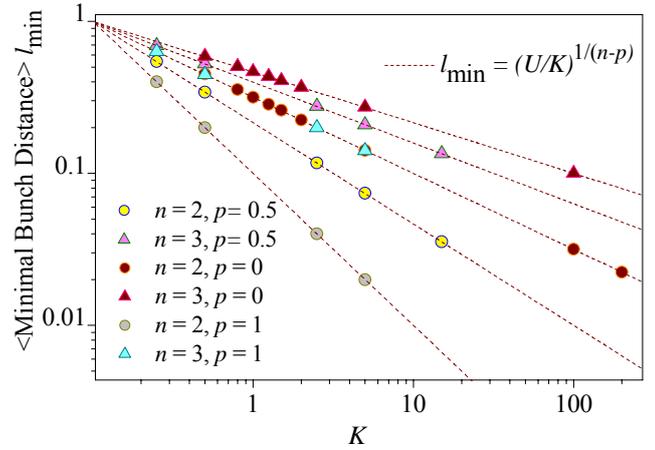

Figure 8 LW2, size dependence of the minimal interstep distance in the bunch. It is not function of the number of steps in the bunch $N$. parameters: $p = 0.5$, $K = 0.5$, $U = 0.1$

Figure 9 LW2, the values of $l_{min}$ when changing the strength of the interstep attraction $K$ and the exponents n and $p$, $U = 0.1$

**Figure 10**

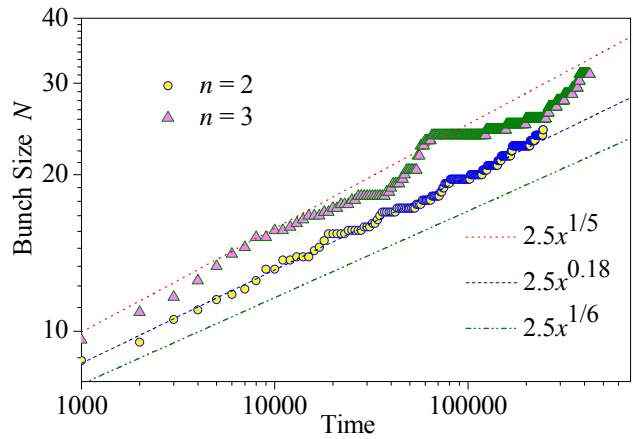

Figure 10 LW2, time scaling of the bunch size $N$. In this case the best fitting exponent for the bunches obtained ($n = 2$) is ~0.18, parameters: $p = 0.5$, $K = 0.5$, $U = 0.1$

6